\documentclass{ws-ijmpcs}

\begin{document}

\markboth{Wim de Boer}
{Borghini's  Mechanism for Dynamic Polarization in Polarized Targets}

%
\catchline{}{}{}{}{}
%

\title{Borghini's  Mechanism for Dynamic Polarization in Polarized Targets}

\author{Wim de Boer}

\address{Karlsruhe Institute of Technology, Physics Department, IEKP, Campus S\"ud, \\Postfach 6980,
76049 Karlsruhe, Germany\\wim.de.boer@kit.edu}

\maketitle

\begin{history}
\received{Day Month Year}
\revised{Day Month Year}
\published{Day Month Year}
\end{history}

\begin{abstract}
This paper is a  contribution to the memorial session for Michel Borghini at the Spin 2014 conference in Bejing, honoring his pivotal role for the development  of polarized targets in high energy physics. 
Borghini proposed for the first time the correct mechanism for dynamic polarization in polarized targets using  organic materials doped with free radicals. In these amorphous materials the spin levels are broadened by spin-spin interactions and g-factor anisotropy, which allows a high dynamic polarization of nuclei by cooling of the spin-spin interaction reservoir. In this contribution I summarize the experimental evidence for this mechanism.  These pertinent experiments were done at CERN in the years 1971 - 1974, when  I was a graduate student under the guidance of Michel Borghini. I finish by shortly describing how  Borghini's spin temperature theory  is now applied  in cancer therapy.
\keywords{dynamc polarization; spin temperature theory; cancer therapy.}
\end{abstract}

\ccode{PACS numbers:}

\section{Introduction}	
Studying the effect of spin in particle interactions has been a topic of interest in high energy physics (see e.g. Ref. \cite{Milner:2013aua} for a review), which required polarized particles, either as target or as beam or both.  The development of polarized targets  at CERN was driven by Michel Borghini, while Alan Krisch from the University of Michigan (Ann Arbor) pushed the polarized beams at Argonne and later at Brookhaven and other accelerators \cite{Krisch:2010hr}. I was lucky enough to work with both of them. After finishing my  Master thesis at the Technical University of Delft on studying spin systems in LMN \cite{deBeer:1973fq}, a material used initially for polarized targets,  I came to CERN as a fellow  in Michel Borghini's group and contributed heavily to  the  experiments leading to the acceptance of Borghini's mechanism  of dynamic nuclear polarization (DNP) in organic materials \cite{Borghini:68,Borghini:1971zza} by ''Dynamic Orientation of Nuclei by Cooling of the Electron Spin-Spin Interactions''\footnote{Michel called this the ''DONKEY" effect, but the name did not stick.}. Here the spin-spin  interactions (SS) comprise all the non-Zeeman energies, which can broaden the Zeeman levels of the free electrons beyond the nuclear Zeeman levels, thus allowing a thermal contact between the SS-reservoir and the nuclear Zeeman reservoir by   electron spin flips in combination with nuclear spin flips. Such a thermal contact is driven at low temperatures mainly by induced spin flips from the polarizing RF field. This dual role of the external RF field (cooling and establishing thermal contact) at low temperatures and in high magnetic fields  was the main new idea from Michel, since DNP by cooling of the spin-spin interactions had been demonstrated before in 1963 by Goldman and Landesman \cite{goldman} in the group of A. Abragam, the world leading expert on DNP at Saclay. but its application to amorphous  materials at low temperatures was far from clear. 

\begin{figure}
\begin{center}
\includegraphics[width=0.48\textwidth]{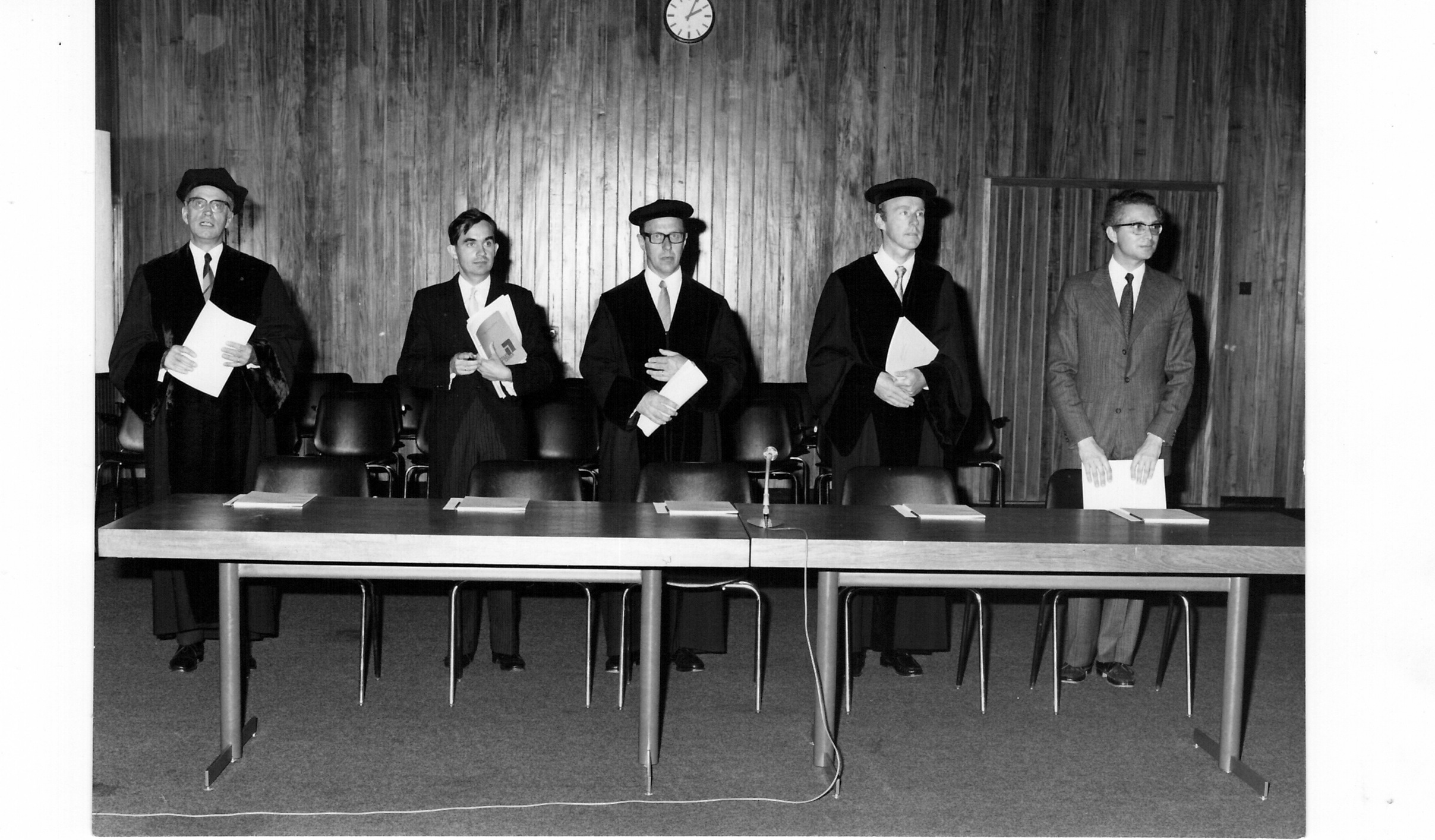}\hspace*{3mm}
\includegraphics[width=0.48\textwidth]{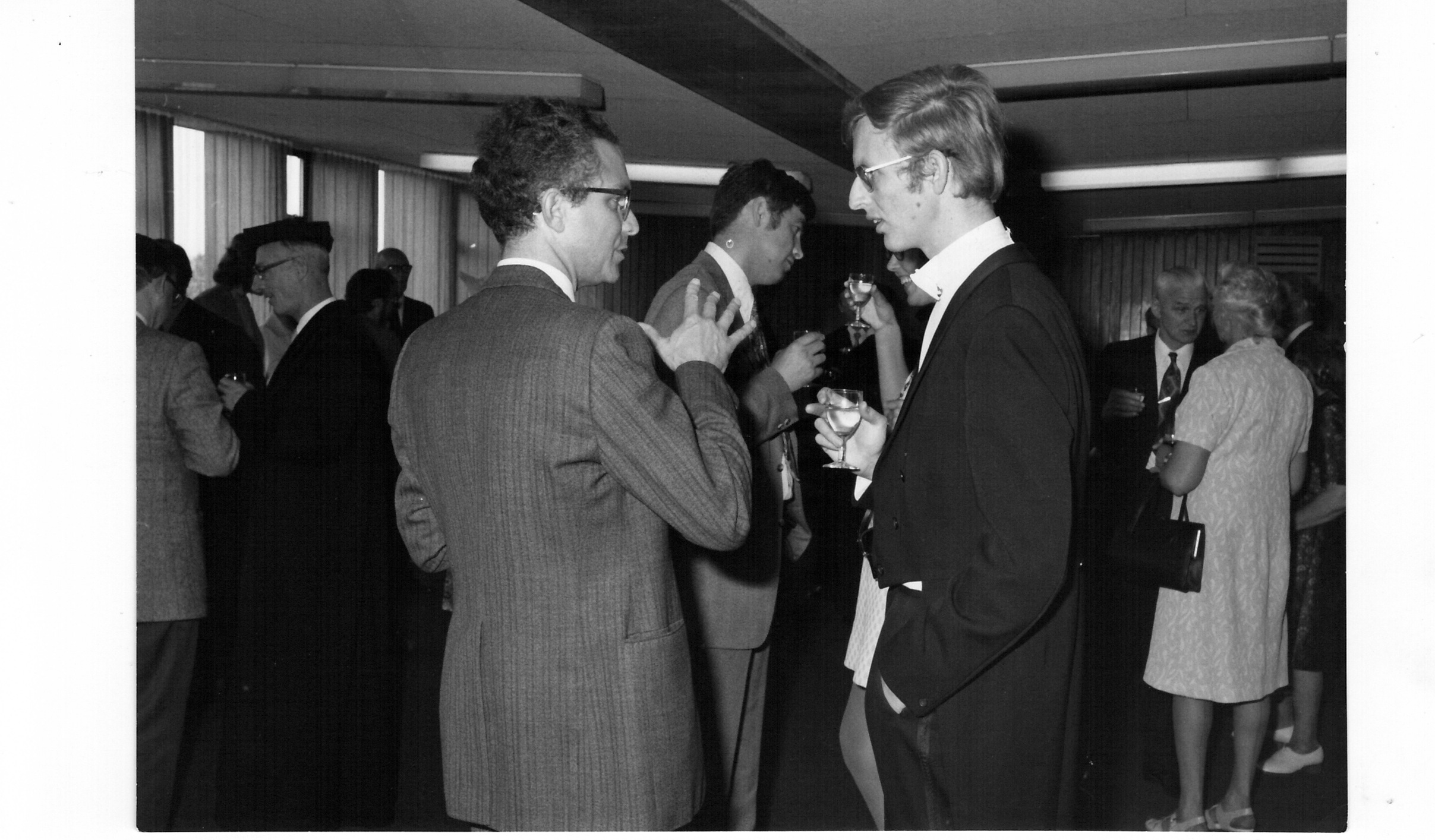}
\caption{a) My thesis Committee at the Technical University of Delft (1974) with Michel on the right. b) Discussion with Michel after the thesis exam.}
\label{f1}
\end{center}
\end{figure}

Borghini, also working in Abragam's group, wrote down his ideas in an extensive thesis. However, his thesis was not accepted by Abragam for reasons unkown to me, but presumably because it lacked experimental verification.
Michel's proposed mechanism was clearly proven by our experiments at CERN, done at temperatures down to 0.1K and magnetic fields up to 5 T.  After all our  results were published by 1976\cite{DeBoer:1972gf,DeBoer:1973fm,deBoer:1973ft,Borghini:1974ch,deBoer:1975pa,deBoer:1975hx,deBoer:1976hx},  Abragam and Goldman wrote a review on DNP, describing in detail our results and recognizing that this was the mechanism of DNP in polarized targets \cite{abragam}. 
These papers were the basis of my PhD thesis \cite{deboerthesis} at the Technical University of Delft. Promotor was Prof. B.S. Blaisse and Michel was a member of the thesis committee, as shown in Fig. \ref{f1}.

When I came to Borghini's group in 1971,  scattering experiments with polarized butanol targets \cite{Mango:1969ww} were in full swing. However, a higher  proton polarization was requested and we continued to work on Michel's  list of possible materials, which should be tried. This was extremely tedious, since every material could be doped with every free radical in a  range of concentrations. Michel left every one much freedom in trying out ideas and organizing his work. This  fosters the creativity of the individual much stronger than in a hierarchical group structure, where everyone is told what to do.   I have kept this working style in my  working groups. We rarely had group meetings, but Michel regularly informed himself how things were going and took care that the infrastructure was optimal, so we had an outstanding mechanical workshop with Jean-Michel Rieubland as driving force behind the actual building and running of the polarized targets, George Gattone as head of the chemistry laboratory, Fred Udo and Huib Ponssen, also two dutch staff members, providing the electronics and digital readout of the polarization. It included for every target an HP2100 computer, which allowed not only a precision determination of the polarization by averaging the sometimes tiny thermal equilibrium signals, but I could also do all calculations for my thesis on ´´my´´ personal computer. The programs were punched with a Teletype writer on paper rolls, which in turn could be read by optical readers. All this was high tech at that time. Then there was of course Tapio Niinikoski, the cryogenic genius, who obtained his thesis on the development of the horizontal dilution refrigerators in the group of Prof. O.V. Lounasma at the Helsinki University of Technology, roughly at the same time as I received my Ph.D thesis  at the Technical University of Delft. Tapio was the mastermind behind the frozen spin targets \cite{Niinikoski:1976jw} and became  the head of the polarized target group  after Michel took over other responsibilities at CERN. 
After getting to know all the tricky details of how to build and operate a polarized target, I was hired by Alan Krisch at the University of  Michigan in Ann Arbor, who had just started to do experiments with polarized beams at Argonne National Laboratory and installed a polarized target. This enabled us to measure the cross section  for a polarized proton beam on a polarized proton target. Surprisingly, this led to  significant spin effects, even  in the total elastic pp cross section for protons with spins parallel or antiparallel, an experimental result \cite{deBoer:1975eb}, which still lacks an interpretation in the frame work of QCD.  But for me the main result was, that high energy physics is at least as interesting as solid state physics, so I became a particle physicist\footnote{The possible unification of all forces in Supersymmetry \cite{Amaldi:1991cn} spurred my interest in dark matter \cite{deBoer:1994dg}, so I later joined,  in addition, the astroparticle physics community in search for the elusive dark matter. }.
In this contribution I want to summarize the exciting experiments done between 1971 and 1974 in Borghini's group at CERN.  
\section{The Theory of Dynamic Polarization}
In  solid {\it crystalline} materials doped with a small concentration of paramagnetic centers with an unpaired free electron the mechanism of dynamic polarization is easy to understand: in a magnetic field H at a temperature T  the relative fraction  of the free electrons  $n_i$ over the spin states with energy $E_i$ is given by the Boltzmann distribution $n_i=exp(-E_i /kT_S)$ for a spin temperature $T_S$. This leads for a spin 1/2 to a polarization $P=(n_+ - n_-)/(n_-+n_-)=\tanh (h\nu /2kT_S)$, where $h$ is Planck's constant and $\nu$ the Larmor frequency of the spin system.  
At high temperatures one can expand the exponential expressions, in which case  the electron and nuclear polarization  are related to the  inverse spin temperature $\beta=h /kT_S$ by $P_e=-\beta\nu_e/2$ and $P_n=\beta\nu_n/2$.  Numerical results for the Larmor frequency  and a spin temperature $T_S$ equal to the lattice temperature lead to an electron polarization $P_e=-0.9975$   in a magnetic field of 2.5 T and a temperature T of 0.5 K, while  the proton polarization $P_n=+0.00511$ under the same conditions. DNP is the art of transferring the high electron polarization to the nuclei via microwave induced spin flips.

For {\it crystalline}  materials the dominant DNP mechanism is the  ''solid'' effect (also called solid-state effect), which was proposed by Abragam and Proctor and verified experimentally, see the review  \cite{abragam} for original references.  In this case one stimulates by microwave irradiation the ''forbidden'' transitions, in which case an electron and neighboring nucleus simultaneously change their spin orientation (either flip-flip or flip-flop transitions, where a flip (flop) indicates a transition to a higher (lower) Zeeman level).
The electron will return to the ground state quickly with a time constant given by the short electron spin-lattice relaxation time of the order of ms. The nucleus has a much longer spin lattice relaxation time, so it does not quickly return to the ground state, but instead it can transfer its polarization to neighboring nuclei via flip-flop spin transitions. This leads to spin diffusion, which  is fast, since energy and angular momentum are conserved. The electron is now ready to polarize the neighboring nucleus again, if it is still receiving photons with the correct energy from an external microwave field. This combination of an external microwave field polarizing neighboring nuclei combined with fast nuclear spin diffusion allows to effectively transfer the high polarization from the electrons to the nuclei.
\begin{figure}
\begin{center}
\includegraphics[width=0.75\textwidth]{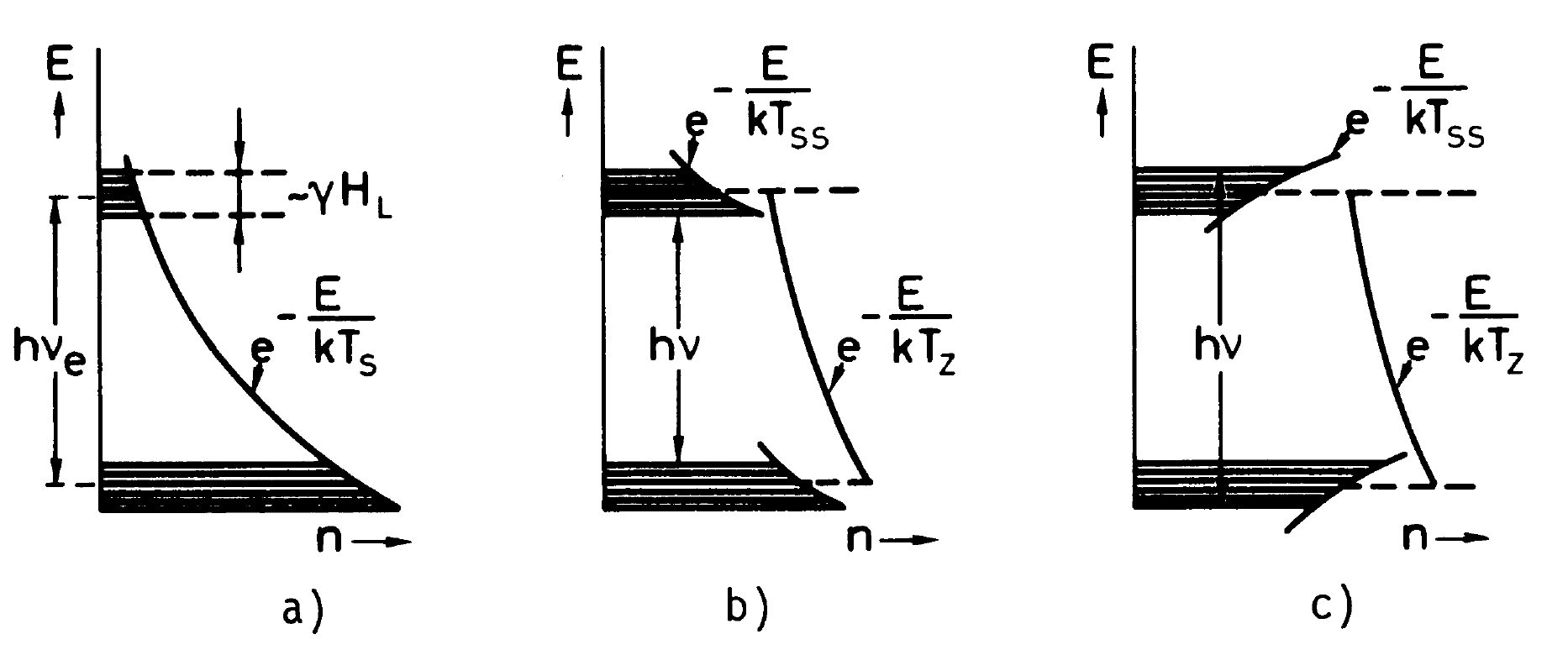}
\caption{a) The equilibrium spin temperature $T_S$ determines the population of the energy levels in a magnetic field. Off-resonance microwave radiation below (above) the Larmor frequency populates the lower (upper) levels inside the broadened Zeeman levels, thus cooling (heating) the SS-reservoir, as indicated in  b) (c). This leads to different temperatures of the Zeeman - and SS-reservoir, indicated by $T_Z$ and $T_SS$, respectively. }
\label{f2}
\end{center}
\end{figure}

In   {\it non-crystalline} solids the spin levels are usually broadened by the different orientations of the molecules, which  experience different internal magnetic fields and this broading  is usually larger than the Zeeman splitting of the nuclei. In this case the resolved solid-state effect will not work anymore, since one is stimulating simultaneously flip-flop and flip-flip transitions\footnote{The net polarization is then given by the difference in intensity of the flip-flop and flip-flip transitions, which is proportional to the difference in intensity of the electron spin resonance line shape and is called the differential solid effect. It always leads  to a nuclear  polarization well below the electron polarization.}.
\begin{figure}
\begin{center}
\includegraphics[width=0.75\textwidth]{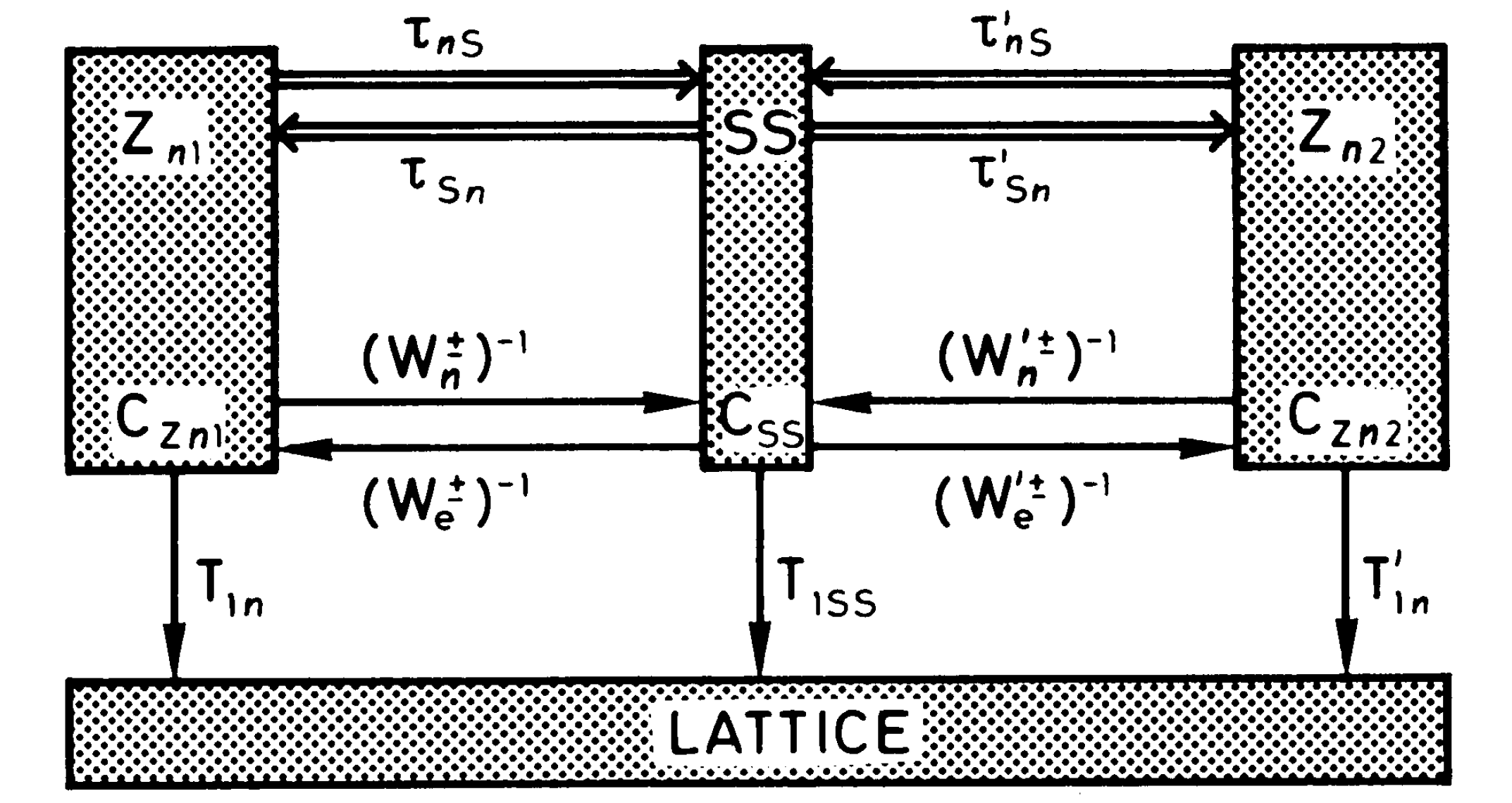}
\caption{Schematic diagram of the thermal contact between two nuclear Zeeman reservoirs  $Z_{ni}$ with heat capacities $C_{Zni}$, the spin-spin interaction reservoir $SS$ and the lattice. The double arrows indicate the thermal contact via the flip-flop transitions between two electrons accompanied with  a nuclear spin flip, while the single arrows indicate the microwave induced ´´forbidden´´ transitions of a simultaneous electron and nuclear spin transition.}
\label{f3}
\end{center}
\end{figure}
However,  another mechanism of DNP may become effective, which is most easily explained by first introducing the concept of a spin temperature  and  a spin-spin interaction (SS) reservoir \cite{goldman,Solomon:63}. These concepts are visualized in Fig. \ref{f2}. In a) the populations of the energy le\-vels follow the Boltzmann distribution, both, for the large Zeeman splitting of the electrons and inside an energy band for a given Zeeman level. However, inside a band, whose width is determined by the non-Zeeman interactions, like the g-factor anisotropy  or spin-spin interactions (SS), the population can be changed by external photons, if one irradiates with   microwave frequencies slightly different from the central Zeeman frequency. This can either cool (Fig. \ref{f2}b) or heat (Fig. \ref{f2}c) the SS-reservoir \cite{redfield,Provotorov:61} and lead even to the highest levels having  the highest population, as shown in Fig. \ref{f2}c, which corresponds to negative spin temperatures of the SS-reservoir. The question is: how strong is the thermal contact between the nuclear Zeeman energy reservoir and this SS-reservoir? This contact can be established either by (i) spontaneous electron spin flip-flops between the Zeeman levels with a simultaneous nuclear spin flip (so a 3-spin process) or this contact can be established by (ii) the microwave induced  forbidden transitions of the  ''solid'' effect.  The different transitions for the thermal contact are schematically indicated in Fig. \ref{f3}. The thermal contact via (i) was demonstrated by Goldman and Landesman \cite{goldman}, who first cooled the SS-reservoir by off-resonance RF irradiation. They then  switched on a magnetic field, which revealed a nuclear polarization, obviously obtained from the thermal contact with the SS-reservoir. However, this method  is unlikely to function well at low temperatures, since then all electron spins are in the lowest state, so there will be few double spin flips between the Zeeman levels of the electrons. Here came the excellent idea of Borghini \cite{Borghini:68,Borghini:1971zza}: he realized that the second method of a thermal contact is independent of the temperature, so it will be the dominant method at low temperature. 
So he extended the Provotorov rate equations  \cite{Provotorov:62} to include the nuclei and solved the three coupled differential equations for the temperatures of the  SS- and Zeeman reservoirs of electrons and nuclei.
The master equation was  well explained by Borghini in his rejected thesis and  I repeated the proof in the appendix of my thesis \cite{deboerthesis}.
However, the formulae were written in the high temperature appro\-ximation, i.e. expanding the exponential function in the Boltzmann distribution. I extended the differential rate equations from Provotorov to low temperatures. The solutions   could still  be written analytically, but they were most easily solved numerically.
Given that we obtained  spin temperatures as low as a few  $\mu K$ , the precision had to be better than 10$^{-10}$, which I could nicely do on "my" HP2100.
\section{Verifying the Mechanism of DNP in Polarized Targets}
\begin{figure}
\begin{center}
\includegraphics[width=0.45\textwidth,height=0.3\textwidth]{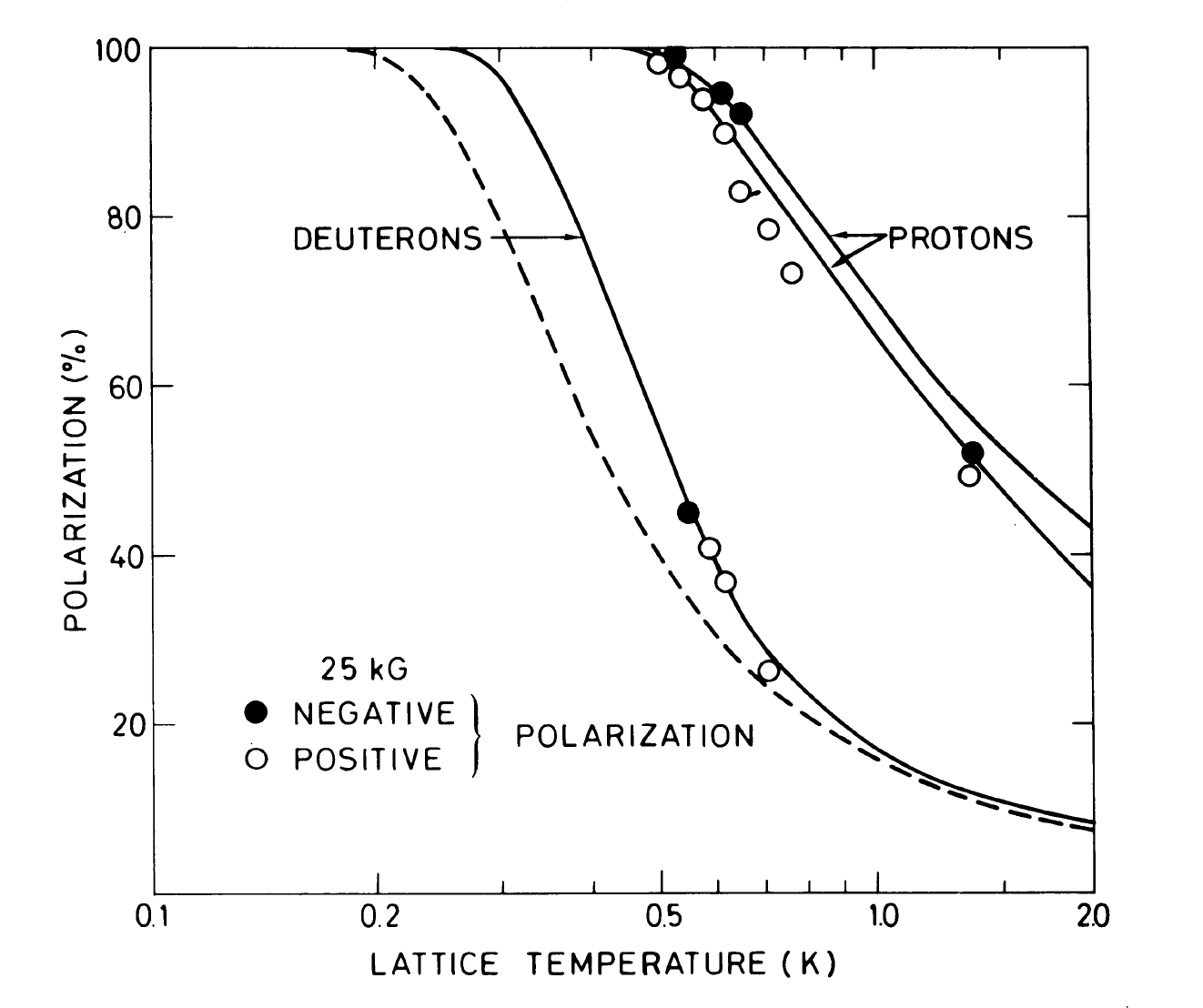}\hspace*{6mm}
\includegraphics[width=0.45\textwidth,height=0.3\textwidth]{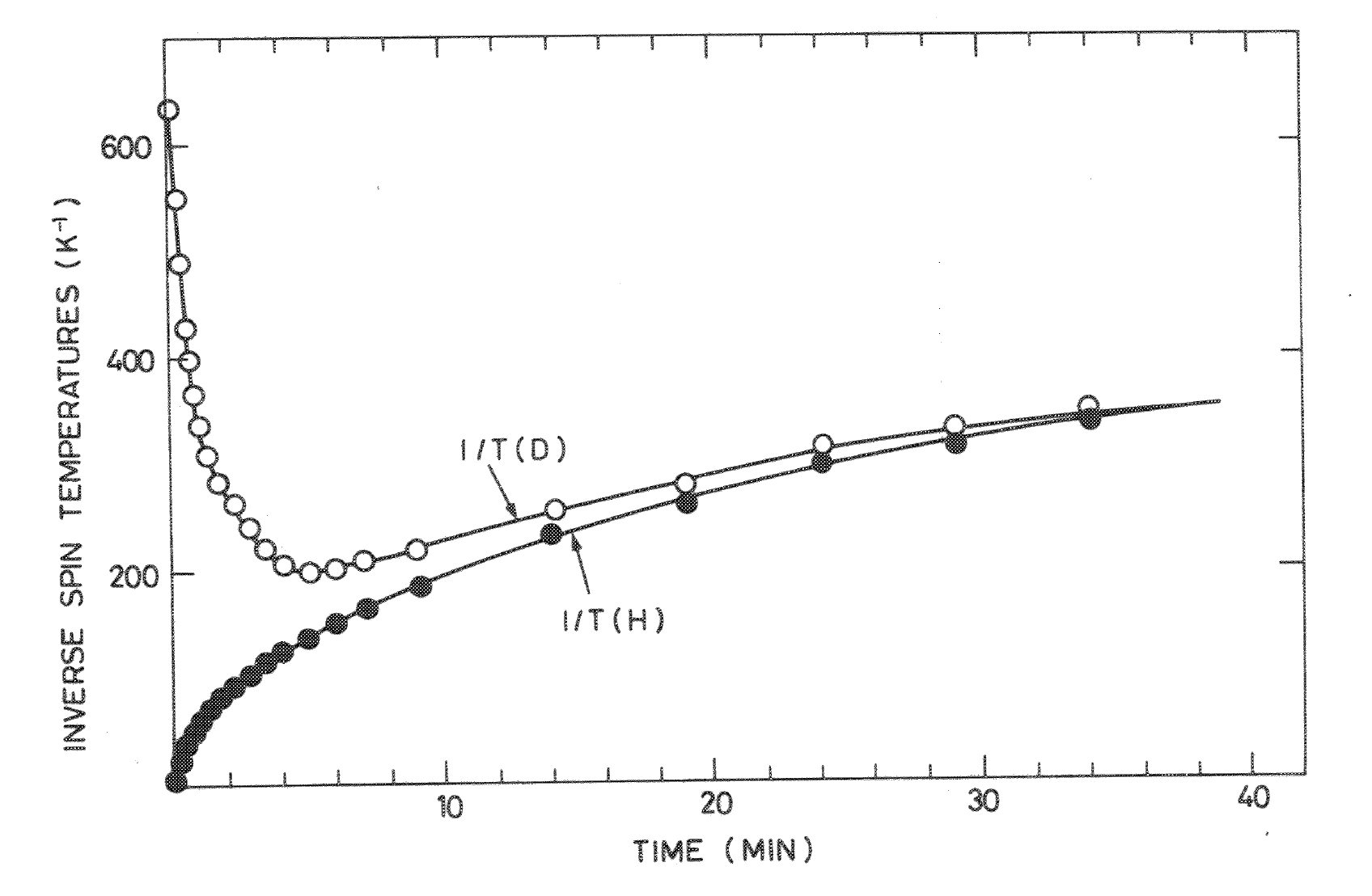}
\center{\hspace*{10mm}a)\hspace*{0.5\textwidth}b)}
\caption{ a) The observed and calculated polarization of protons and deuterons as function of the lattice temperature. b) Evolution of the spin temperature of protons and deuterons after destroying the proton polarization and switching on a microwave field to polarize again. The rapid equalization of the spin temperatures shows the increase of the thermal contact via  the forbidden transitions (single arrows in Fig. \ref{f3}). They finally reach the common temperature $T_{SS}$ of the spin-spin interaction reservoir.}
\label{f4a}
\end{center}
\end{figure}

The first polarized targets consisted of frozen butanol beads doped with a free radical and reached a  proton polarizations of about 40\%  \cite{Mango:1969ww}. A few years later propanediol doped with Cr-V complexes were used, in which a proton polarization close to 100\% was obtained \cite{DeBoer:1972gf,DeBoer:1973fm}. Such a high polarization would be impossible for the differential solid-state effect. So only the cooling via the SS-reservoir remained a possibility and we started a program to prove this.
\begin{figure}
\begin{center}
\includegraphics[width=0.45\textwidth,height=0.3\textwidth]{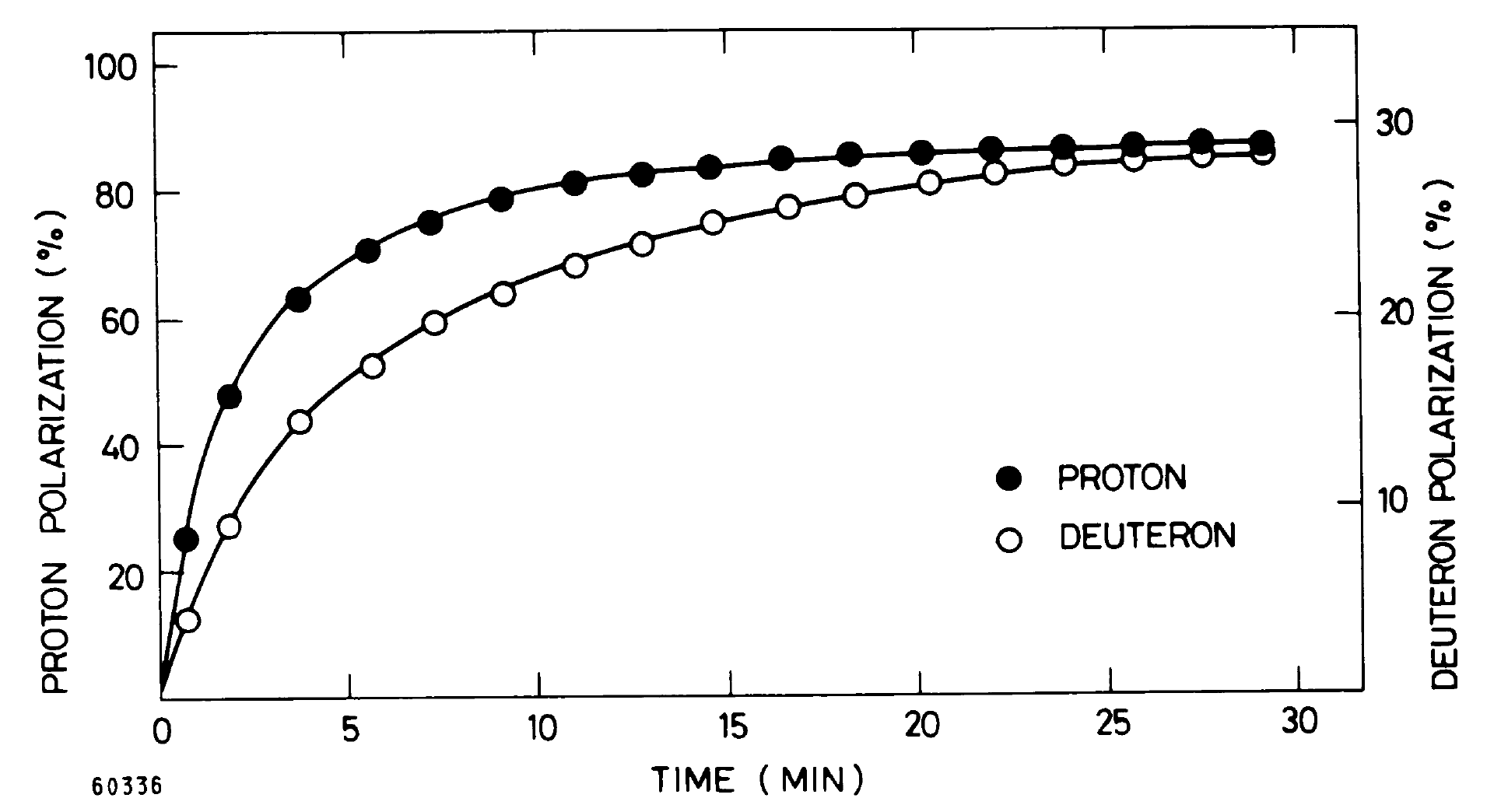}\hspace*{3mm}
\includegraphics[width=0.45\textwidth,height=0.315\textwidth]{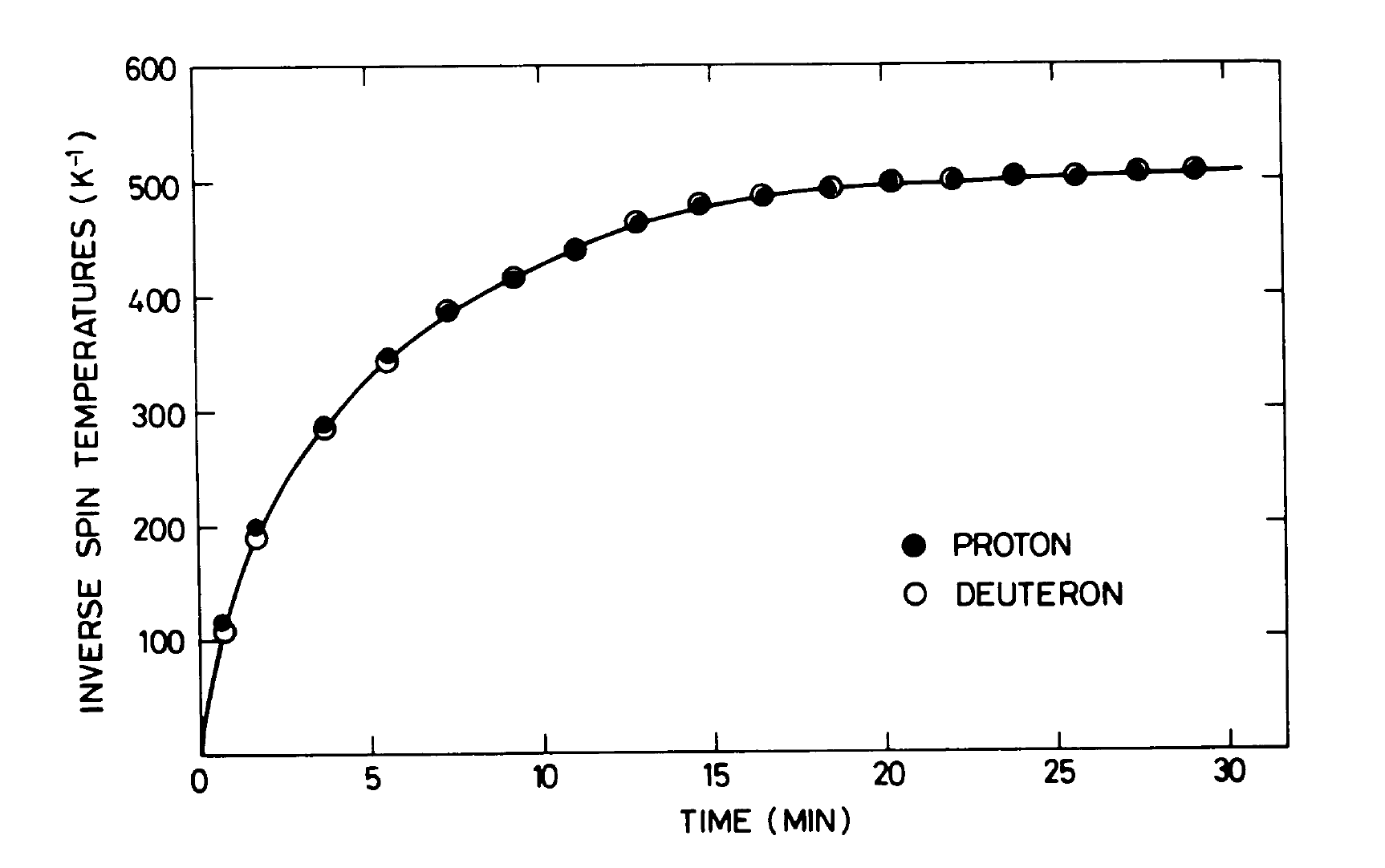}
\center{\hspace*{10mm}a)\hspace*{0.5\textwidth}b)}
\caption{a) Build-up of the proton and deuteron polarization (indicated on the left- and right-hand scale, resp.) in a partially deuterated sample. b) Build-up of the spin temperature of protons and deuterons.}
\label{f4}
\end{center}
\end{figure}
The predictions of  Borghini's spin temperature model are clear: 
several nuclear species with different Larmor frequencies obtain a different polarization, but they have the same spin temperature,  if the thermal contact is good enough, where good enough means that the leakage to the lattice is small in comparison with the heat transfer between the reservoirs in Fig. \ref{f3}. 
This could be verified  by observing the proton and deuteron polarization in a partially deuterated propanediol sample, as demonstrated in Fig. \ref{f4}. During the polarization build-up the polarization becomes different (left side), but the spin temperature of the two nuclear species stays the same (right side). 
In the thermodynamical model of Fig  \ref{f3} the final polarization depends on the leakage to the lattice, which is a strong function of  temperature \cite{DeBoer:1973fm}.
Fig. \ref{f4a}a shows the predicted and observed polarization of protons and deuterons in propanediol doped with Cr-V complexes as function of temperature \cite{deBoer:1976hx}. Satisfactory agreement between theory and experiment is obtained. Here the heat capacities of the nuclear Zeeman reservoirs and temperature dependence of the spin-lattice relaxation times were carefully taken into account. The effect of the reduced heat capacity of the deuteron system is clearly seen by the difference between the dashed and solid line for the deuterons.
At low temperatures the nuclear Zeeman reservoirs  in Fig. \ref{f3}  are rather isolated, at least without microwave irradiation establishing a thermal contact (single arrows in Fig. \ref{f3}).  This allows to destroy the polarization of one nuclear species by inducing spin transitions between the lower and upper nuclear Zeeman levels with a saturating RF field. If one e.g. destroys the proton polarization in a highly polarized sample, the deuteron polarization stays high. After switching on the microwave irradiation to polarize the sample again, this microwave irradiation establishes the thermal contact between the protons and deuterons, thus equalizing their spin temperatures much faster than expected from the polarization time by cooling of the SS reservoir. This is demonstrated in Fig. \ref{f4a}b, which clearly proves the dual role of the microwave field at low temperatures, the original idea of Michel.

Since the deuteron Larmor frequency is  smaller than the width of the proton Zeeman levels, one has the same situation between deuterons and protons as for protons and electrons. Therefore, one should be able to polarize the deuterons by off-center irradiation of a polarized proton spin system, thus cooling the SS-reservoir of the protons.
Several schemes are possible: first polarize a sample of protons and deuterons by cooling the SS-reservoir of the electrons via off-center irradiation of the electron Zeeman levels with microwaves. Then destroy  the deuteron polarization with a saturating RF field, followed by a cooling of the proton SS-reservoir with another RF field close to the proton Larmor frequency.  Since the deuteron has S=1 the various levels can be populated such, that a pure tensor polarization can be obtained. Many experiments have been done and they all confirm the spin temperature theory in a quantitative way \cite{deBoer:1973ft,deBoer:1975pa}. Because of lack of space these beautiful experiments will not be described here. 
\section{Experiments showing different  Mechanisms of DNP}
\begin{figure}
\begin{center}
\includegraphics[width=0.31\textwidth,height=0.3\textwidth]{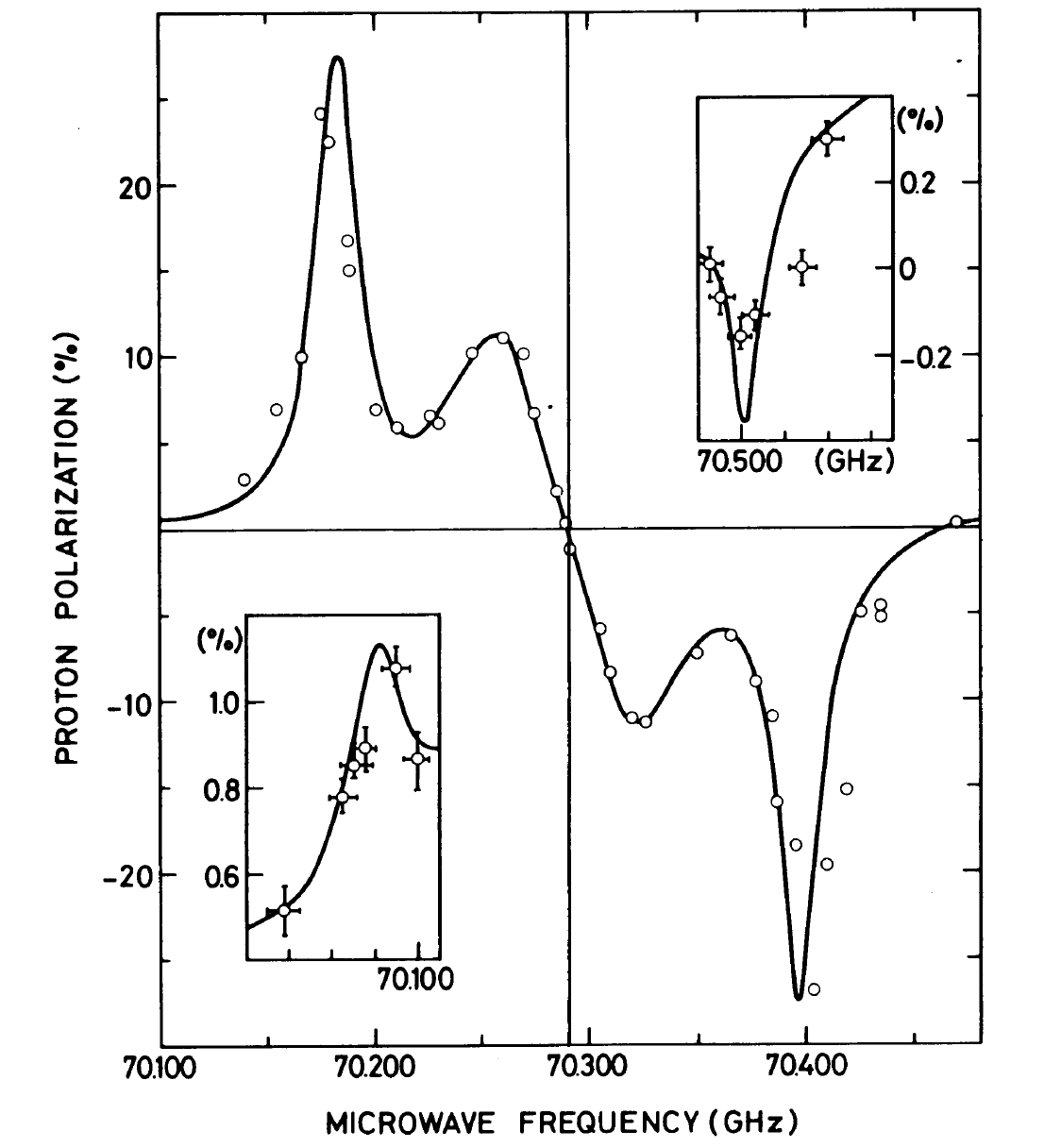}\hspace*{3mm}
\includegraphics[width=0.3\textwidth,height=0.31\textwidth]{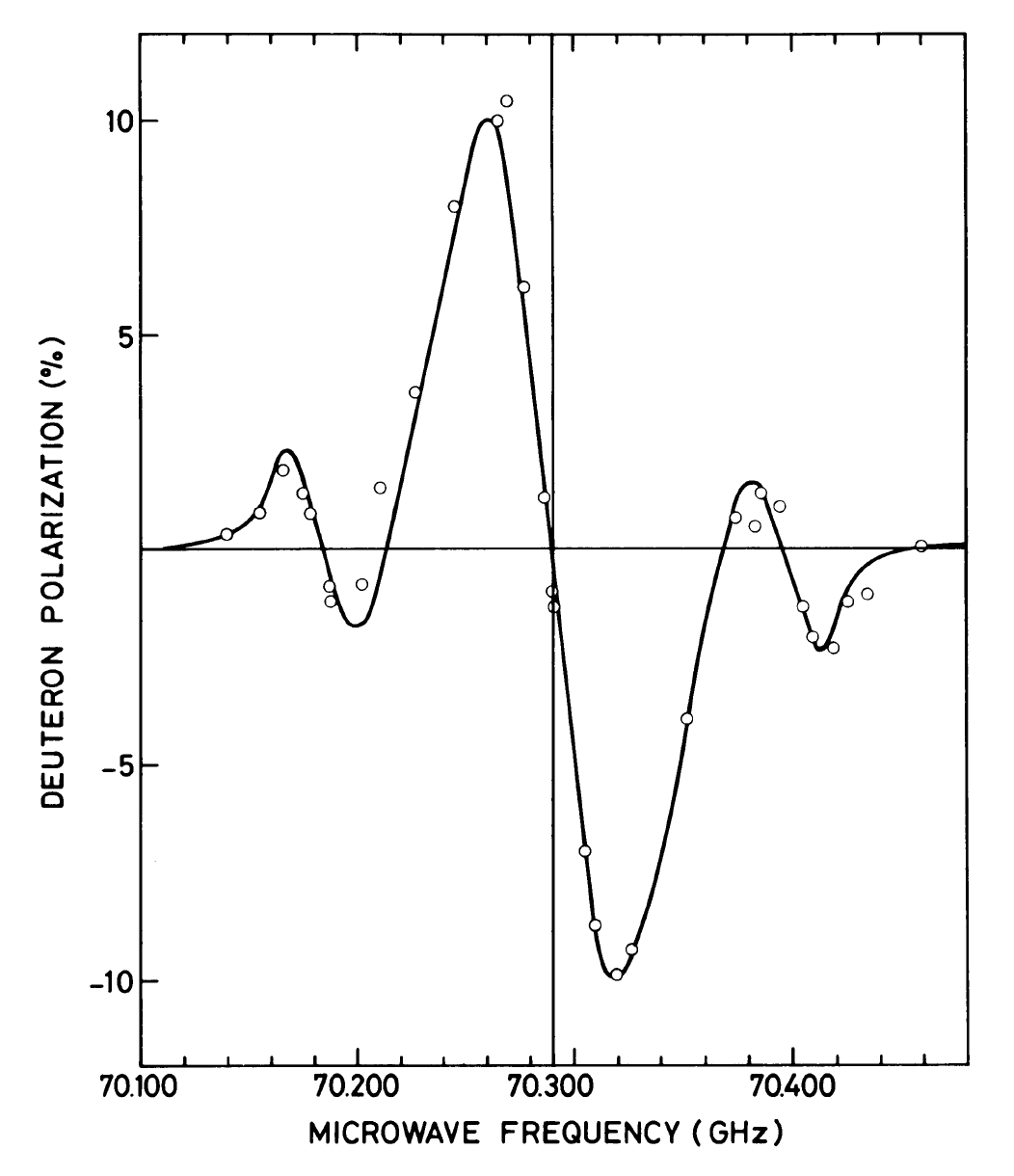}\hspace*{3mm}
\includegraphics[width=0.3\textwidth,height=0.32\textwidth]{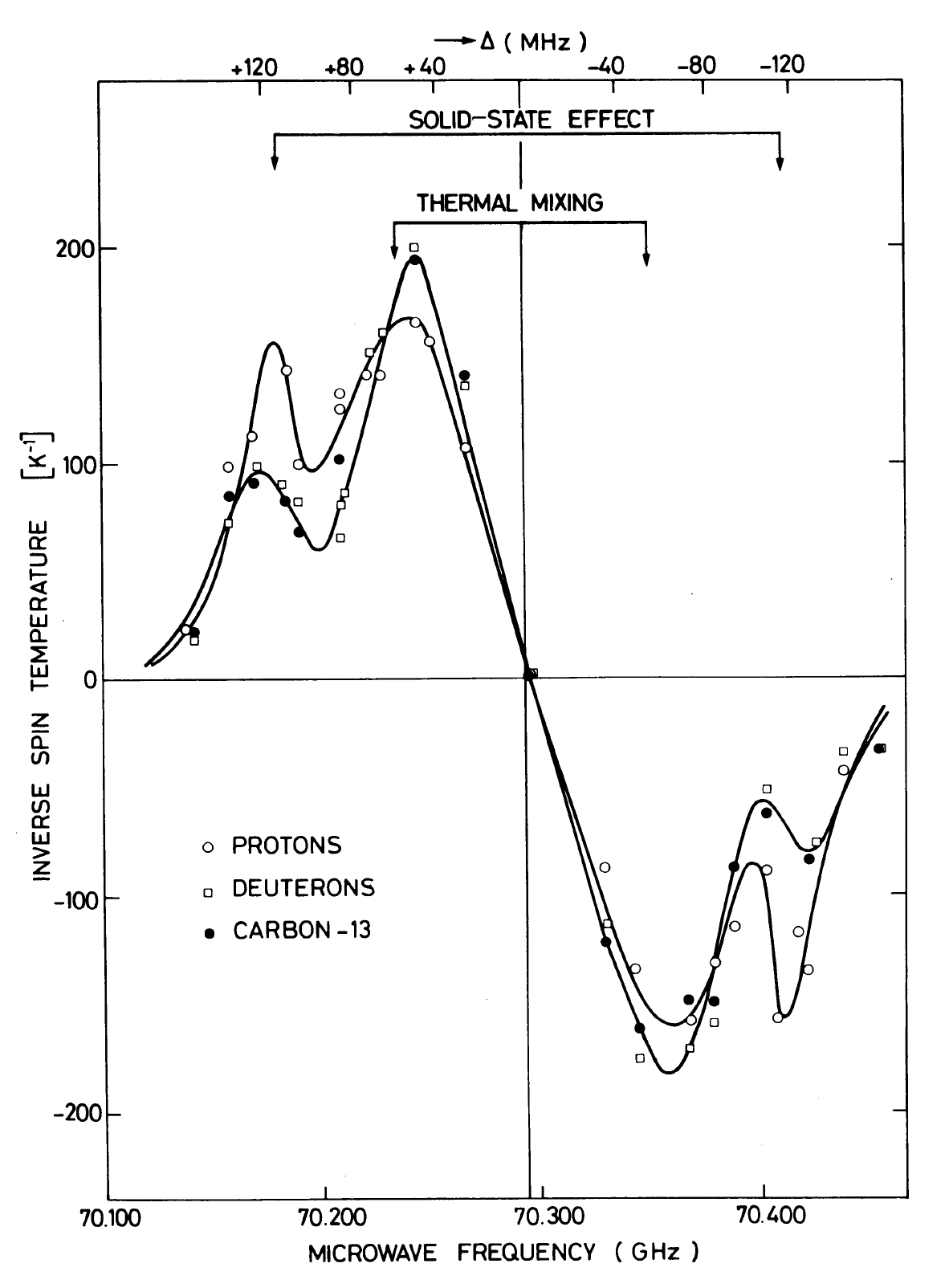}
\center{\hspace*{10mm}a)\hspace*{0.3\textwidth}b)\hspace*{0.3\textwidth}c)}
\caption{a) Proton polarization as function of microwave frequency. b) Deuteron polarization as function of microwave frequency. c) Inverse spin temperature as function of microwave frequency.}
\label{f5}
\end{center}
\end{figure}

As mentioned before, the solid-state mechanism of DNP is effective for narrow Zeeman energy levels of the unpaired electrons of the free radicals, while the mechanism via cooling of the spin-spin interaction reservoir is effective for electron Zeeman le\-vels broader than the nuclear Zeeman splitting. The free radical BDPA has a width of the Zeeman levels, which is below the nuclear Zeeman levels of protons, but above the ones for deuterons and $^{13}$C nuclei. Therefore one expects  a combination of the two mechanisms, which should occur at different microwave frequencies. This was indeed the case as proven in a frozen sample of  partially deuterated m-xylene (2,2-D6) doped with BDPA ($6\cdot10^{18}$ spins/cm$^3$).  The measurements were performed in a magnetic field of 2.5T at a temperature of 0.75K \cite{deBoer:1975hx,deBoer:1976hx,deboerthesis}.
The proton polarization is shown as function of microwave frequency in Fig. \ref{f5}a. The inner peaks correspond to the polarization by the cooling of the SS-reservoir, while the outer peaks at frequencies $\nu_e\pm\nu_p$ correspond to the solid-state effect. The insets show the double solid-state effect at frequencies $\nu_e\pm 2\nu_p$ corresponding to a simultaneous spin flip of an electron and two protons.
Deuterons have a  Larmor frequency below the width of the electron Zeeman levels, so the polarization by the forbidden transitions of an electron and deuteron spin flip are not visible, since they are too close to the frequencies of the optimum cooling of the electron SS-reservoir. However, the double solid state effect of a simultaneous triple spin flip of an electron, proton and deuteron at frequencies  $\nu_e\pm \nu_p \pm \nu_D$ leads to the four peaks in the deuteron polarization outside the main peak from the cooling of the SS-reservoir in Fig. \ref{f5}b.
The maximum polarization of protons and deuterons by  cooling of the SS-reservoir (inner peaks in Figs. \ref{f5}a and b) is about 10\%. From a comparison with Fig. \ref{f4} it is clear that this does not correspond to an equal spin temperature, presumably because of the poor thermal contact between  the proton Zeeman reservoir and the SS-reservoir. A good thermal contact requires the width of the electron Zeeman levels to be large with the nuclear Zeeman splitting, which is the case for the deuteron system, but not for the proton spin system. To check this hypothesis we prepared a sample of toluol-D8 with a larger concentration of BPA ($5\cdot10^{19}$ spins/cm$^3$), which increases the width of the electron Zeeman levels. In addition, we measured the polarization of the $^{13}C$ nuclei, which have a small Zeeman splitting as well and hence, should obtain the same spin temperature as the deuterons. This is indeed the case, as shown in Fig. \ref{f5}c. The protons obtain indeed  almost the same spin temperature with this sample with an order of magnitude higher concentration of BPDA leading to a broadening of the electron Zeeman levels.
\section{Application of DNP in cancer therapy}
Dynamic polarization has found an actively pursued application  in cancer research:  polarized $^{13}$C nuclei in tracers of tumors yield a strongly enhanced signal in Magnetic    Resonance  Imaging   (MRI), so  smaller tumors  can be discovered, see Ref. \cite{Rodriques:2014} for a recent development and references therein. The medical people call this hyperpolarization, but the polarization happens in setups similar to the ones used in polarized targets, see e.g. \cite{Eichhorn:2013}.
The surprising discovery: after thawing the samples in a magnetic field, the polarization is largely maintained in the gas phase. Dissolving the gas into a liquid and injecting it into the body yields strongly enhanced NMR signals of the tumors. The polarization lasts only minutes, but this is enough for a picture in a modern magnetic resonance tomograph.

\section{Summary}
The mechanisms of dynamic polarization in polarized targets are by now well understood. According to Michel Borghini's idea this happens via a two-step process between the different heat reservoirs:  i) cooling or heating of the SS-reservoir by off-center microwave irradiation; ii) establishing thermal contact between the SS-reservoir and the nuclear Zeeman reservoirs by the same microwave irradiation inducing triple spin-flips, namely a flip-flop transition of two electrons combined with a nuclear spin transition (either up or down until thermal equilibrium is reached). This mechanism was proven by many different experiments showing the thermal contact between the different reservoirs. By extending the usual high temperature approximation to low temperatures the spin temperature theory was proven to be valid to spin temperatures in the $\mu$K range, as was evident from the excellent agreement between theory and experiment. Nowadays the polarized targets, invented for high energy experiments, are used to enhance the polarization in biological tracers used to find tumors in Magnetic Resonance Imaging. The enhanced polarization provides a strongly enhanced signal, thus allowing to detect smaller tumors. The relatively high $^{13}$C polarization of up to 60\% is a clear manifestation, that Borghini's proposed mechanism of the dynamic polarization by cooling of the electron spin-spin interaction reservoir is at work.
Michel certainly would have been delighted to see that his idea of dynamic polarization has found such important applications in fields never thought of before. 


\begin{thebibliography}{10}

\bibitem{Milner:2013aua}
R.~G. Milner, {\em PoS} {\bf PSTP2013}, p.~3  (2013).

\bibitem{Krisch:2010hr}
A.~Krisch, {\em 13th Workshop on High Energy Spin Physics (DSPIN-09), Dubna,
  Russia, C09-09-01}   (2010).

\bibitem{deBeer:1973fq}
R.~de~Beer, W.~de~Boer, C.~van~'t Hof and D.~van Ormondt, {\em Acta
  Crystallogr.B Struct.Crystallogr.Cryst.Chem.} {\bf 29}, 1473  (1973).

\bibitem{Borghini:68}
M.~Borghini, {\em Phys.Lett.} {\bf 26A}, p. 242  (1968).

\bibitem{Borghini:1971zza}
M.~Borghini, {\em Proc. 2nd International Conference on Polarized Targets,
  Berkeley, CA, USA} (Ed. G. Shapiro) {\bf C710830}, 1  (1971).

\bibitem{goldman}
M.~Goldman and A.~Landesman, {\em Phys.Rev.} {\bf D132}, p. 610  (1963).

\bibitem{DeBoer:1972gf}
W.~de~Boer, {\em Nucl.Instrum.Meth.} {\bf 107}, 99  (1973).

\bibitem{DeBoer:1973fm}
W.~de~Boer and T.~Niinikoski, {\em Nucl.Instrum.Meth.} {\bf 114}, 495  (1974).

\bibitem{deBoer:1973ft}
W.~de~Boer, M.~Borghini, K.~Morimoto, T.~Niinikoski and F.~Udo, {\em
  Phys.Lett.} {\bf B46}, 143  (1973).

\bibitem{Borghini:1974ch}
M.~Borghini, W.~de~Boer and K.~Morimoto, {\em Phys.Lett.} {\bf A48}, p. 244
  (1974).

\bibitem{deBoer:1975pa}
W.~de~Boer, {\em Phys.Rev.} {\bf B12}, 828  (1975).

\bibitem{deBoer:1975hx}
M.~Borghini, W.~de~Boer and K.~Morimoto, {\em Phys.Lett.} {\bf A48}, p. 244
  (1974).

\bibitem{deBoer:1976hx}
W.~de~Boer, {\em J.Low.Temp.Phys.} {\bf 22}, p. 185  (1976).

\bibitem{abragam}
A.~Abragam and M.~Goldman, {\em Rep. Prog. Phys.} {\bf 41}, p. 395  (1978).

\bibitem{deboerthesis}
W.~de~Boer, {\em PhD thesis, Tech. Univ. of Delft,} {\bf CERN Yellow Report,
  CERN-74-11}  (1974), {http://cds.cern.ch/record/186203/files/CERN-74-11.pdf}.

\bibitem{Mango:1969ww}
S.~Mango, O.~Runolfsson and M.~Borghini, {\em Nucl.Instrum.Meth.} {\bf 72}, 45
  (1969).

\bibitem{Niinikoski:1976jw}
T.~Niinikoski and F.~Udo, {\em Nucl.Instrum.Meth.} {\bf 134}, p. 219  (1976).

\bibitem{deBoer:1975eb}
W.~de~Boer, R.~C. Fernow, A.~Krisch, H.~Miettinen, T.~Mulera {\em et~al.}, {\em
  Phys.Rev.Lett.} {\bf 34}, 558  (1975).

\bibitem{Amaldi:1991cn}
U.~Amaldi, W.~de~Boer and H.~F\''urstenau, {\em Phys.Lett.} {\bf B260}, 447
  (1991).

\bibitem{deBoer:1994dg}
W.~de~Boer, {\em Prog.Part.Nucl.Phys.} {\bf 33}, 201  (1994).

\bibitem{Solomon:63}
I.~Solomon, {\em Proc. Magnetic and Electric Resonance and Relaxation,
  Amsterdam, Ed. J. Schmidt} , p.~25  (1963).

\bibitem{redfield}
A.~Redfield, {\em Phys.Rev.} {\bf 98}, p. 1787  (1955).

\bibitem{Provotorov:61}
B.~N. Provotorov, {\em JETP} {\bf 14}, p. 1126  (1961).

\bibitem{Provotorov:62}
B.~N. Provotorov, {\em JETP} {\bf 15}, p. 611  (1962).

\bibitem{Rodriques:2014}
T.~Rodrigues, E.~Serrao, B.~W.~C. Kennedy, D.~Hu, K.~M. I. and K.~Brindle, {\em
  Nature Medicine} {\bf 20}, p. 93â97  (2014).

\bibitem{Eichhorn:2013}
T.~Eichhorn, Y.~Takado, N.~Salameh et al., {\em Proc. of the Nat. Acad. of
  Science of the USA} {\bf 110(45)}, 18064  (2013).

\end{thebibliography}

\end{document}